\documentclass[twocolumn,floatfix,superscriptaddress,amsmath,showpacs,showkeys,aps,pre]{revtex4-2}

\usepackage{t1enc}
\usepackage[final]{graphicx}
\usepackage{graphicx}
\usepackage{epsfig}
\usepackage{bm}
\usepackage{color}
\usepackage[normalem]{ulem}
\usepackage{graphics}

\begin{document}
\title{Tricritical behavior in a neural model with excitatory and inhibitory units}

\author{Joaquin Almeira}

\affiliation{Instituto de F\'isica Enrique Gaviola (IFEG-CONICET) Ciudad Universitaria, 5000 C\'ordoba, Argentina}

\author{Tomas S. Grigera}
\affiliation{Instituto de F\'isica de L\'iquidos y Sistemas Biol\'ogicos (IFLYSIB), CONICET y Universidad Nacional de La Plata, La Plata, Argentina}
\affiliation{Departamento de F\'isica, Facultad de Ciencias Exactas, Universidad Nacional de La Plata, La Plata, Argentina}
\affiliation{Istituto dei Sistemi Complessi, Consiglio Nazionale delle Ricerche, via dei Taurini 19, 00185 Rome, Italy}
 \affiliation{Consejo Nacional de Investigaciones Cient\'{\i}fcas y Tecnol\'ogicas (CONICET), Buenos Aires, Argentina}
 
\author{Dante R. Chialvo}
\affiliation{Instituto de Ciencias F\'isicas (ICIFI-CONICET), Center for Complex Systems and Brain Sciences (CEMSC3), Escuela de Ciencia y Tecnolog\'ia, Universidad Nacional de Gral. San Mart\'in, Campus Miguelete, San Mart\'in, Buenos Aires, Argentina}
\affiliation{Consejo Nacional de Investigaciones Cient\'{\i}fcas y Tecnol\'ogicas (CONICET), Buenos Aires, Argentina}

\author{Sergio A.\ Cannas}
\affiliation{Instituto de F\'isica Enrique Gaviola (IFEG-CONICET) Ciudad Universitaria, 5000 C\'ordoba, Argentina}

\affiliation{Facultad de Matem\'atica Astronom\'ia F\'isica y Computaci\'on, Universidad Nacional de C\'ordoba.}
\affiliation{Consejo Nacional de Investigaciones Cient\'{\i}fcas y Tecnol\'ogicas (CONICET), Buenos Aires, Argentina}

 \date{\today}

\begin{abstract}
While the support for the relevance of critical dynamics to brain function is increasing, there is much less agreement on the exact nature of the advocated critical point.  Thus, a considerable number of theoretical efforts are currently concentrated on which mechanisms and what type/s of transition can be exhibited by neuronal networks models.  In that direction, the present work describes the effect of incorporating a fraction of inhibitory neurons on the collective dynamics.  As we show, this results in the appearence of a tricritical point for highly connected networks and non-zero fraction of inhibitory neurons.  We discuss the relation of the present results with relevant experimental evidence.
\end{abstract}

%\pacs{}
%\keywords{}

\maketitle

\section{Introduction}

 A large repertoire of diverse spatiotemporal activity patterns in the brain is the basis for adaptive behaviour.
Understanding the manner in which the brain is able to form and reconfigure a large range of cortical configurations, in a flexible manner, remains an unsolved challenge.  A leading proposal interprets this large repertoire as the expected, generic, large diversity of states near instabilities, which is composed, by its own nature, of a mixture of ordered and disordered patterns.  In more technical terms, near the critical point of a second order phase transition it is known that the system exhibits the largest number of metastable states which is only limted by the system size. In the brain, these metastable states would correspond to cortical configurations, or patterns of activation. In a nutshell, this is the basis of the ``brain criticality hypothesis'',  suggested as the solution to the above mentioned challenge \cite{beggs,chialvo2010,mora}. In that regards, several key experimental works have demonstrated, over the last decade, that brain dynamics at large and small scales meets the requirements of critical dynamics, including finite-size scaling of the correlation length \cite{fraiman,Haimovici2013,ribeiro}, power-law distribution of activation clusters \cite{tagliazucchi1}, and dynamic scaling \cite{camargo} among the most significant findings. 

Despite these advances, the exact nature of the advocated critical point is not fully understood yet.  Thus, most theoretical efforts are currently concentrated on what type/s of transition can be exhibited by neuronal networks models as well as in searching for falsifiable predictions able to identify the correct model. 
In that direction, the present work generalizes the results previously described \cite{Zarepour} in a neuronal model with excitatory interactions running on a Watts-Strogatz topology.  By investigating the effects of adding inhibitory interactions we uncover the presence of a tricritical point for  a non-zero fraction of inhibitory neurons, in the regime of high connectivity.  The paper is organized as follows: In section II we describe the model as well as the simulation and the finite-size scaling methods used. The results are presented in section III, and the relevance of the main findings is discussed in section IV.

\section{Model and Methods}
\subsection{The model}

In this work we use  a generalization of the neural model presented in Ref.\cite{Zarepour} in which a fraction $f$ of neurons are inhibitory. To this end, we associate a variable $\epsilon_i=\pm 1$ to each neuron $i$, where $\epsilon=-1$ represents an inhibitory neuron and $\epsilon=+1$  an excitatory one. The value of each the variables $\{ \epsilon_i \}$ is chosen independently with probability $f$ to be $\epsilon_i=-1$ and  $1-f$ to be $\epsilon_i=+1$. Those values are kept fixed during the network evolution. The model runs 
over a small-world network with a weighted adjacency matrix $w_{ij}$.  The network topology is obtained  following the usual  Watts-Strogatz recipe\cite{watts-strogatz}. That is, we start from a ring of $N$ nodes in which each node is connected symmetrically to its $2m$ nearest neighbors. Then, for each node each vertex connected to a clockwise neighbor  is rewired to a random node  with a probability $\pi$ and preserved with probability $1-\pi$, so the average degree $\langle k \rangle=2m$ is preserved\cite{Barrat}. This algorithm  provides a non-weighted symmetric adjacency matrix $A_{ij}=A_{ji}=0,1$. Then,  the  weighted adjacency matrix $w_{ij}=w_{ij}$ is obtained by assigning to every non-null link  $A_{ij}\neq 0$  a   random real value  chosen from an exponential distribution $p(w)=\lambda\, e^{-\lambda\, w}$ with $\lambda=12.5$. This procedure mimics the weights distribution of the human connectome\cite{Haimovici2013,Hangmann2008}.

The node dynamics of the neural model responds to  the Greenberg-Hastings cellular automaton\cite{Greenberg-Hastings}, in which each node
 $i$ of the network has associated a three state dynamical  variable $x_i=0,1,2$, corresponding to the following dynamical states: quiescent  ($x_i=0$), excited  ($x_i=1$) and refractory  ($x_i=2$). The transition rules are the following: if a node at the discrete time $t$ is in the quiescent state $x_i(t)=0$ it can make a transition to the excited state $x_i(t+1)=1$ with a small probability $r_1$ or if $\sum_j w_{ji}\, \epsilon_j\, \delta(x_j(t),1) > T$, where $T$ is a threshold and $\delta(x,y)$ is a Kronecker delta function; otherwise, $x_i(t+1)=0$. If it is excited $x_i(t)=1$ then it becomes refractory $x_i(t+1)=2$ always. If it is refractory $x_i(t)=2$ then it becomes quiescent  $x_i(t+1)=0$ with probability $r_2$ and remains refractory $x_i(t+1)=2$  with probability $1-r_2$. Following Refs.\cite{Haimovici2013, Zarepour} we set $r_1=10^{-3}$ and $r_2=0.3$

\subsection{Analysis of the dynamical transition}
\label{statistics}

We focus on dynamical clusters of coherent activity, namely groups of simultaneously activated nodes ($x_i=1$) which are linked through non zero weights $w_{ij}$.  It is known that for $f=0$ the system presents a dynamical phase transition separating a regime where the active clusters are isolated from one where such clusters span across the whole system.  The transition can be continuous or discontinuous depending on the values of the topological parameters $\langle k \rangle$ and $\pi$ \cite{Zarepour,Sanchez}.  As we will show in the next section, varying $f$ can also change the transition from continuous to discontinuous at fixed topological parameters.  Here we explain the metrics used to characterize the transition.

We simulate the model at several values of $\pi$, $\langle k \rangle$, and $f$, and different network sizes $N$. Each simulation is started from a random distribution of activated  sites, and the system is let to run  $500$ time steps before starting data collection.  We found this time interval to be enough for the system to reach a stationary state for any system size and for any value of the network parameters.  We compute several observables to describe a percolation-like transition as a function of $T$ and $f$. Specifically, we calculate the average size of the largest (i.e., giant) cluster, $\langle S_1 \rangle$. For very large systems, this quantity provides the standard percolation order parameter $P_\infty = \lim_{N\to\infty} \langle S_1 \rangle/N$, namely the probability of an arbitrary node to belong to the infinite percolating cluster.  We also compute the average size of the second largest cluster $\langle S_2 \rangle$, together with the average cluster size (or susceptibility),
\begin{equation}\label{averages}
    \langle s \rangle = \frac{\sum'_s  s^2 N_s}{\sum'_s  s N_s}
\end{equation}
where the primed sum runs over all cluster sizes except the giant one and $N_s$ is the number of clusters of size $s$ \cite{Barrat,Margolina}.  We find that on varying the control parameter ($T$ or $f$), $\langle S_1\rangle$ can change from zero to finite both continuously or discontinuously.

When the transition is continuous, we analyze it as in standard percolation.  In this case both $\langle s \rangle$ and   $\langle S_2 \rangle $ are expected to exhibit (size-dependent) maxima for a certain pseudo critical value of the control parameter (the threshold $T$ or the fraction $f$), that scales with system size as \cite{Stauffer} $\langle s \rangle \sim N^{\gamma/\nu d}$, $\overline{S}_2 \sim N^{d_f/d}$. Here $\gamma$ and $\nu$ are the standard susceptibility and correlation length critical exponents, $d$ is the effective dimension of the system and $d_f$ the fractal dimension of the percolating cluster. 

To characterize transition in the discontinuous case we used two different methods:

\emph{(A) Order parameter hysteresis analysis:} \cite{Sanchez}. For fixed values of $N$, $\langle k \rangle$, $\pi$ and $f$, we keep track of $S_1$ as $T$ is slowly increased at a fixed rate from some initial value $T_0$ up to some maximum value $T_F$, and then decreased again down to $T_0$ at the same rate, without resetting the neurons states when changing $T$.  We set the rate of change of the control parameter by changing $T \to T+ \Delta T$ every $t_1$ steps. The values of $T_0$ and $T_F$ were chosen such that the location of the maxima of $\langle s \rangle$ and $S_2$ fall inside the interval $[T_0,T_F]$. As in the $f=0$ case \cite{Sanchez}, we verified in many cases the presence of well-defined hysteresis loops for values of  $T_- < T < T_+$, were the border values $T_\pm$ depend on $\langle k \rangle$, $\pi$ and $f$. In all the simulations we used $\Delta T=5\times 10^{-4}$ and for every set of parameters we performed several checks using values of $t_1$ between $10^2$ and $10^ 4$. If the values $T_\pm$ turned out to be independent of $t_1$ in that range (within errors) the transition temperature was estimated as the average of the hysteresis loop, $T_t = (T_- + T_+)/2$. When the hysteresis loop showed a strong dependency on $t_1$, we switched to the next method to estimate the transition temperature. 

\emph{(B) Order parameter histograms analysis:} For fixed values of $N$, $\langle k \rangle$, $\pi$ and $f$, we computed a histogram of the values of the order parameter ${S}_1$ along a single, long simulation run, for different values of $T$. Close to a discontinuous transition, one expects such distribution to show a two-peak structure for long enough simulation times (i.e., for periods of time such that the system evolution provides a  good sampling of  both phases).  The transition temperature can then be estimated as the value of $T$ for which both peaks are the same height.  This method is useful when the probability of jumping from one phase to the other is relatively high (moderate systems sizes and/or close enough to a critical point), so that the characteristic flip time between phases is small compared with the simulation time. The histogram method is very well established for studying first-order phase transitions in systems under thermodynamic equilibrium \cite{Landau}. The consistency of our results shows that the method  can also work in non equilibrium discontinuous transitions.

%%%%%%%%%%%%%%%%%%%%%%FIGURE 1%%%%%%%%%%%%%%%%%%%%%
\begin{figure}
\begin{center}
\includegraphics[width=0.42\textwidth]{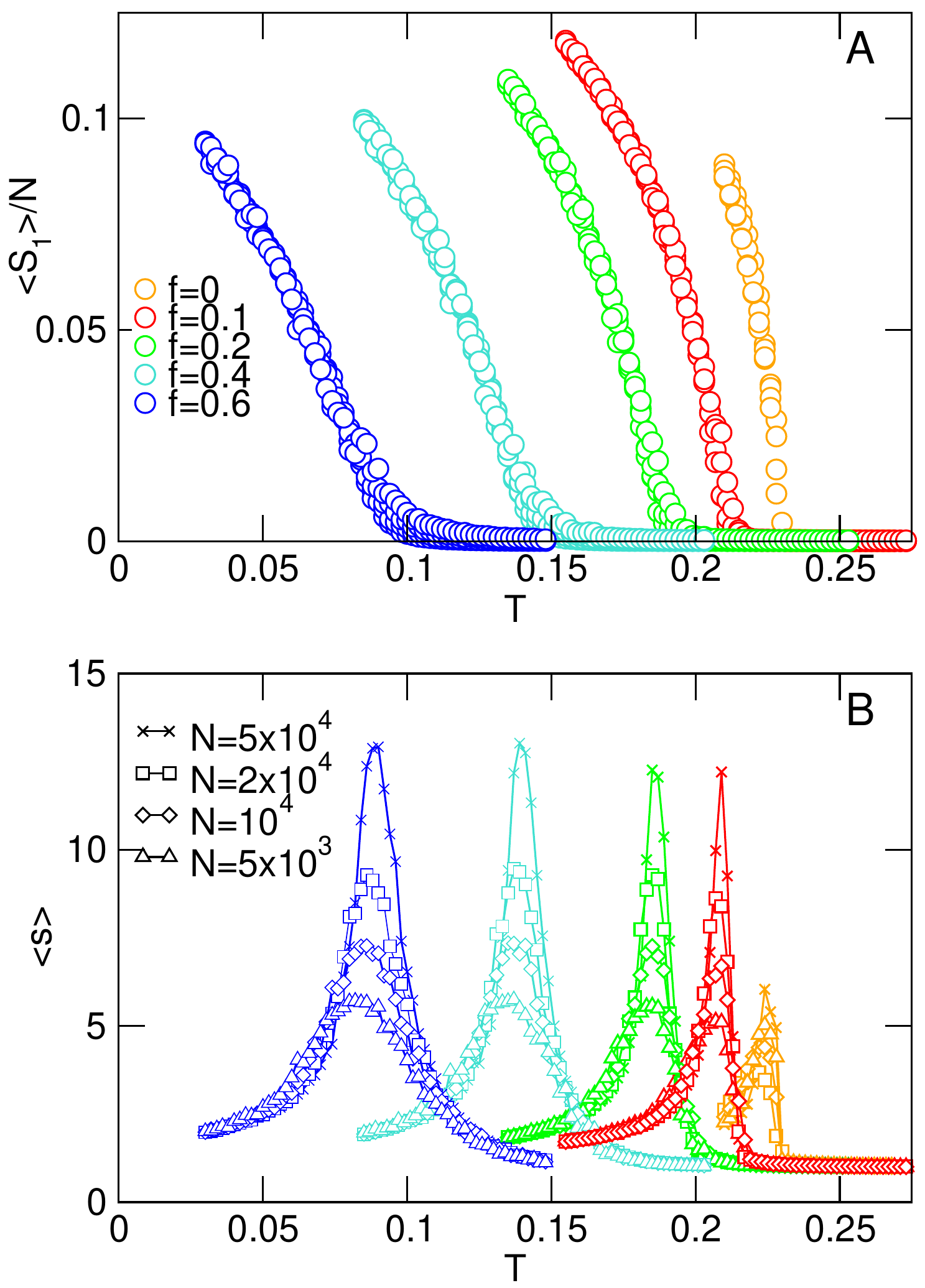}
\caption{\label{fig1} (Color online) Order parameter (panel A) and  corresponding susceptibility (panel B) for   different values of $f$ and system sizes $N$. Different colors denote $f$  values and the different symbols  indicate the system sizes.  Network parameters: $\pi=0.6$ and $\langle k \rangle=16$.}
\end{center}
\label{figure1}
\end{figure}

\section{Results}

For $f=0$, i.e., in the absence of inhibitory neurons, the model corresponds to the case studied in Ref. \cite{Zarepour}. It can exhibit  different dynamical regimes, including a percolation-like phase transition  between high and low activity regimes, depending on the topological parameters of the underlying  network. Such transition can be of second order (i.e., critical) for intermediate values of $\langle k \rangle$ and high enough values of $\pi$, or first-order like (discontinuous) for large enough values of $\langle k \rangle$ \cite{Zarepour}. 

We started our analysis by considering the effect of including inhibitory neurons in a network whose topological parameters correspond to the second-order region for $f=0$.  We found that the presence of inhibitory neurons does not eliminate the continuous transition.  On the contrary, $f$ acts as a new control parameter for the transition, as shown in Fig.\ref{fig1} for $\langle k \rangle=16$ and $\pi=0.6$. In other words, the transition can be observed (i.e., a size dependent maximum of $\langle s \rangle$ at the point where the order parameter almost falls to zero) either by changing $T$ for fixed $f$ (see Fig.\ref{fig1}) or by changing $f$ for fixed $T$ (not shown). Hence, we have  a line of critical points in the $(f,T)$ space, whose universality class will be analyzed later.  Very similar results were obtained for other values of $(\langle k \rangle,\pi)$ in the critical region for $f=0$ (see Fig.4 of Ref.\cite{Zarepour}).

%%%%%%%%%%%%%%%%%%%%%%FIGURE 2%%%%%%%%%%%%%%%%%%%%%
\begin{figure}
\begin{center}
\includegraphics[width=0.42\textwidth]{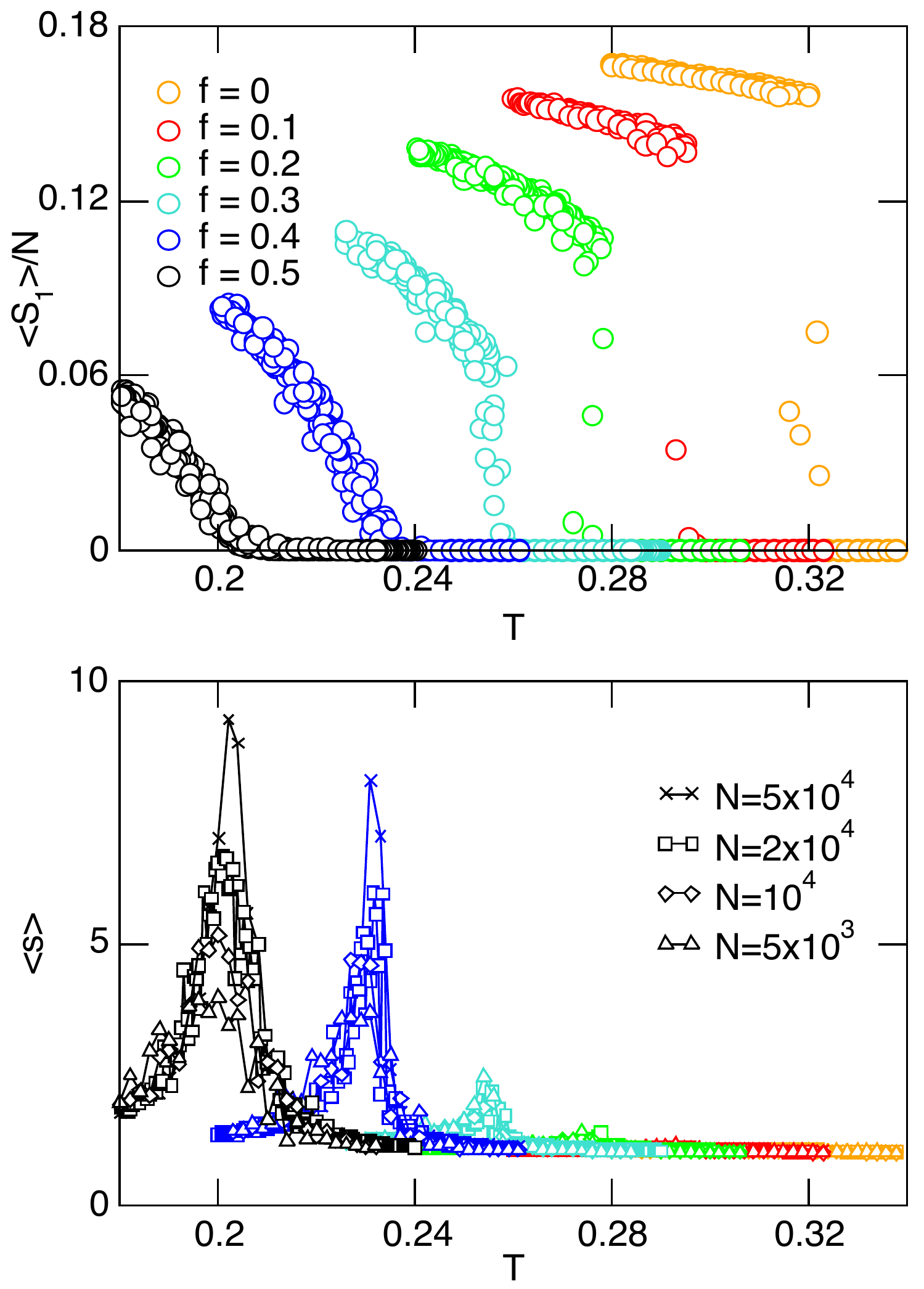}
\caption{\label{fig2} (Color online) Order parameter (panel A) and  corresponding susceptibility (panel B) for different values of $f$ and system sizes $N$. Different colors denote $f$  values and the different symbols  indicate the system sizes.  $\pi=0.6$ and $\langle k \rangle=30$.}
\end{center}
\label{figure2}
\end{figure}

Next, we analyzed the influence of inhibitory neurons on the dynamics when the topological parameters for $f=0$ give rise to a discontinuous transition. The typical behavior of the order parameter and $\langle s \rangle$ as a function of $T$ for fixed $f$ is shown in Fig.\ref{fig2} for $\langle k \rangle=30$ and $\pi=0.6$. We see that for small fractions of inhibitory neurons the transition remains discontinuous, giving rise to a first order transition line. However, as $f$  increases, the nature of the transition changes smoothly to second order,  where the  maximum of $\langle s \rangle$ starts to exhibit a strong size dependency.  This suggests the presence of a tricritical point  where the first and second order transition lines meet. In order the characterize better this phenomenon, we first performed a detailed calculation of the first order transition line, using the two methods described in section \ref{statistics}.

\begin{figure}
\begin{center}
\includegraphics[width=0.42\textwidth]{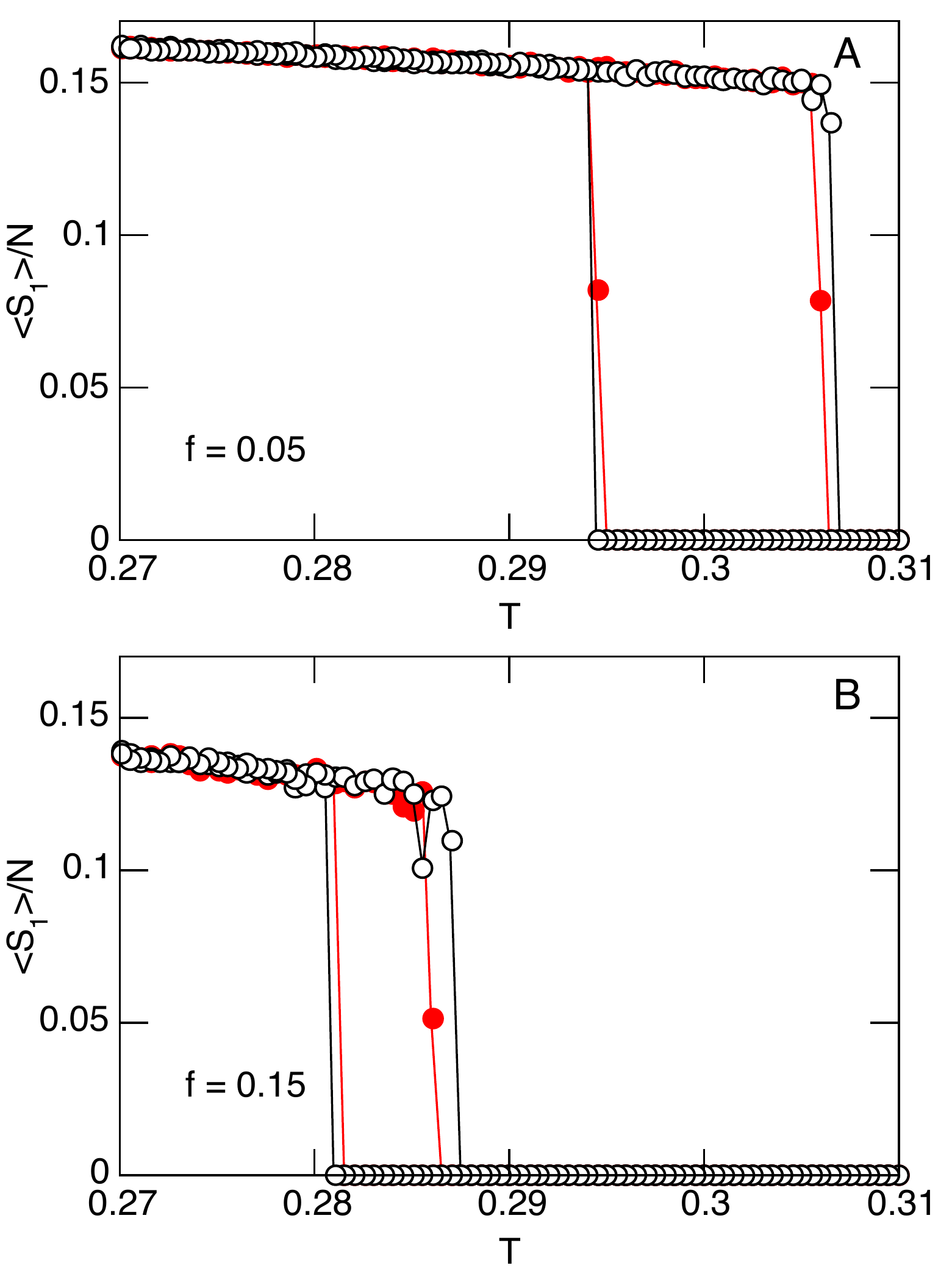}
\caption{\label{fig3} (Color online)  Hysteresis loops of the order parameter and two (fixed) fractions of  inhibitory neurons: $f=0.05$ in panel A and  $f=0.15$ in panel B. The threshold $T$ is changed $T \to T + \Delta T$ with  $\Delta T=5\times 10^{-4}$, every $t_1$ simulation steps. Empty black  symbols correspond to $t_1=100$ and filled red symbols to $t_1=10^4$.  The network parameters are $\pi=0.6$,  $\langle k \rangle=30$ and  $N=2\times 10^4$.  }
\end{center}
\label{figure3}
\end{figure}

\begin{figure}
\begin{center}
\includegraphics[width=0.45\textwidth]{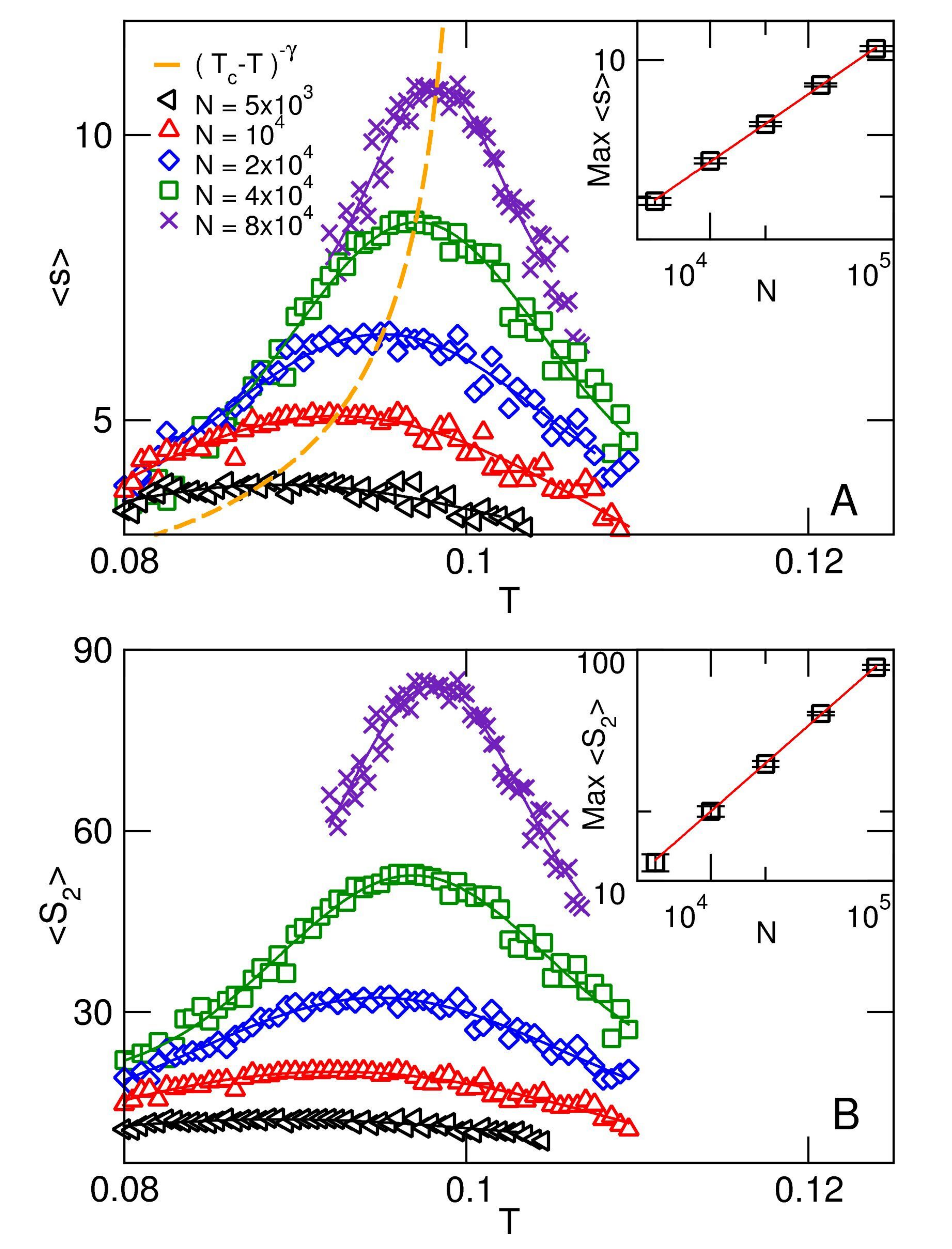}
\caption{\label{fig5} (Color online) Finite size scaling  for   $\langle k \rangle=30$, $\pi=0.6$, $f=0.8$. The insets show the scaling with $N$ of the maxima of the corresponding quantities. Continuous lines  are convenient fitting functions to estimate the maxima. Panel A: $\langle s \rangle$  {\it vs.} $T$. Numerical fitting of the maxima gives $\gamma/\nu d = 0.37 \pm 0.01$. Dashed line corresponds to the scaling relation between $T$ at  the peak susceptibility and system size $N$:  $<s> = A*(T_c - T)^{-\gamma}$. Fitting the location of the $<s>$ peaks, we found $A=0.22, T_c =0.101$ and $\gamma=0.66$. Panel B: $\langle S_2 \rangle$  {\it vs.} $T$.  Numerical fitting of the maxima gives $d_f/ d = 0.70 \pm 0.02$. }
\end{center} 
\end{figure}

Hereafter we will focus on the  $\langle k \rangle=30$ and $\pi=0.6$ case. For small enough values of $f$ (i.e., up to $f\approx 0.25$) we observe well defined hysteresis loops (namely, independent of the rate of change of $T$) as shown in Fig.\ref{fig3}. We also observe that the area of the hysteresis loops shrinks as $f$ increases and tends to disappear for $f \approx 0.3$, giving a first estimation of the tricritical point location. However, a strong dependency on the rate of change of $T$ emerges for $f > 0.25$ and the method looses accuracy in that region.
As we depart from the tricritical point by further increasing $f$,  we can estimate the transition line through the finite-size scaling of the maxima of $\langle s \rangle$ and/or $S_2$, for large enough system sizes. Actually both quantities do not peak at the same value, but the difference becomes negligible for system sizes larger than $N=2\times 10^4$. An example for $f=0.8$ is shown in Fig.\ref{fig5}.
%%%%%%%%%%%%%%%%%%%%%%FIGURE 3%%%%%%%%%%%%%%%%%%%%%

%%%%%%%%%%%%%%%%%%%%%%FIGURE 4%%%%%%%%%%%%%%%%%%%%%
 
\begin{figure}
\begin{center}
\includegraphics[width=0.42\textwidth]{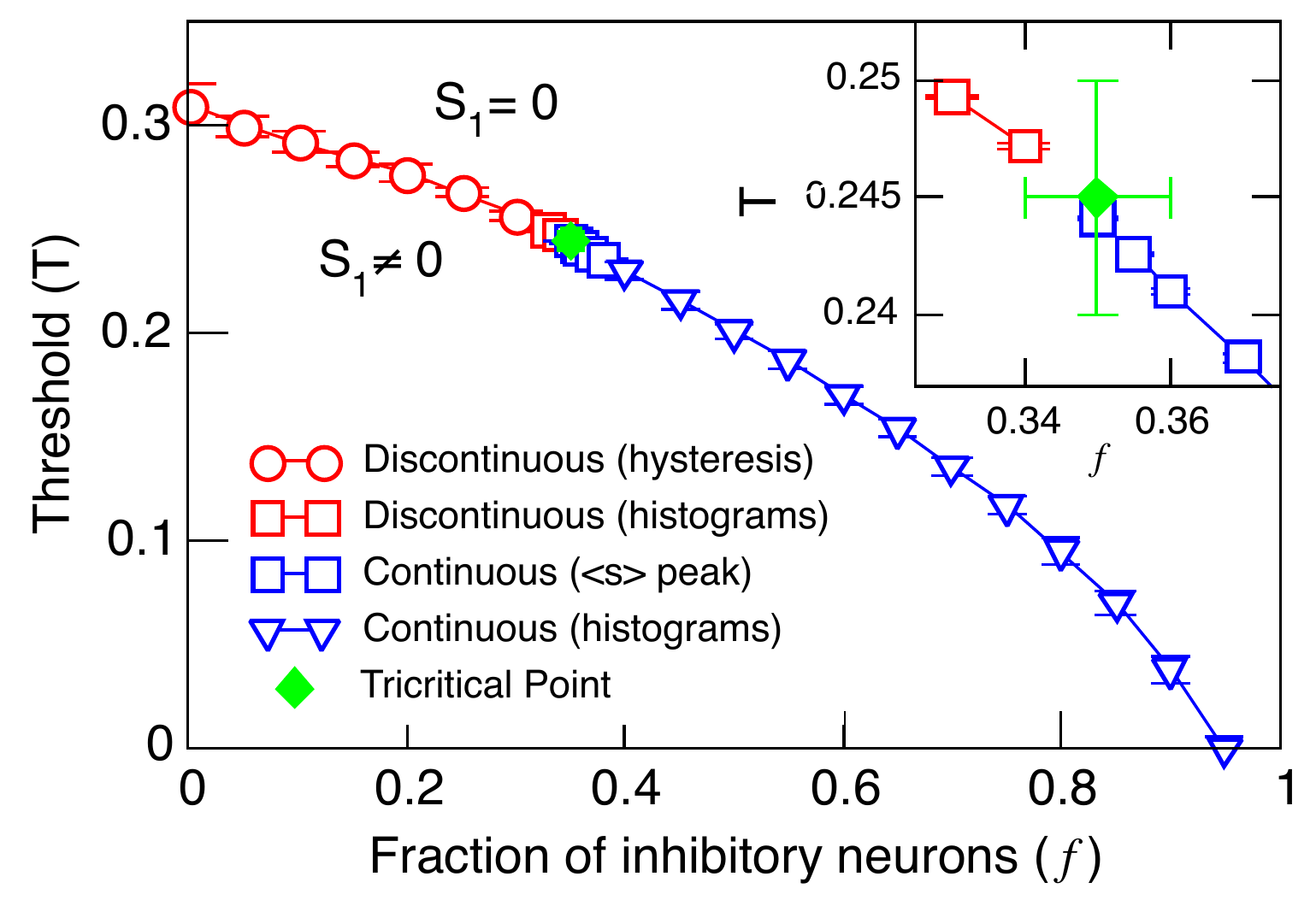}
\caption{\label{fig6} (Color online) Phase diagram in $(f,T)$  space for\\  $\langle k \rangle=30$ and $\pi=0.6$. Blue symbols correspond to the continuous phase transition while red points corresponds to a discontinuous one. Circles correspond to first order transition points estimated by order parameter  hysteresis cycles for $N=2\times 10^4$.  Triangles correspond to second order transition points estimated by the location of the $\langle s \rangle$ peak for $N=2\times 10^4$.  Squares correspond to transition points (both first and second order) obtained through the finite size behavior of the order parameter histograms.}
\end{center}
\end{figure}

We summarize all the previous results in the phase diagram in $(f,T)$  space shown in Fig.\ref{fig6}.  We see that both transition lines (first and second order) meet at the tricritical point (indicated by a star) with equal slope within numerical errors, as expected\cite{Goldenfeld}, showing the consistency of  our original assumption. 
%%%%%%%%%%%%%%%%%%%%%%FIGURE 5%%%%%%%%%%%%%%%%%%%%%

\begin{figure}
\begin{center}
\includegraphics[width=0.42\textwidth]{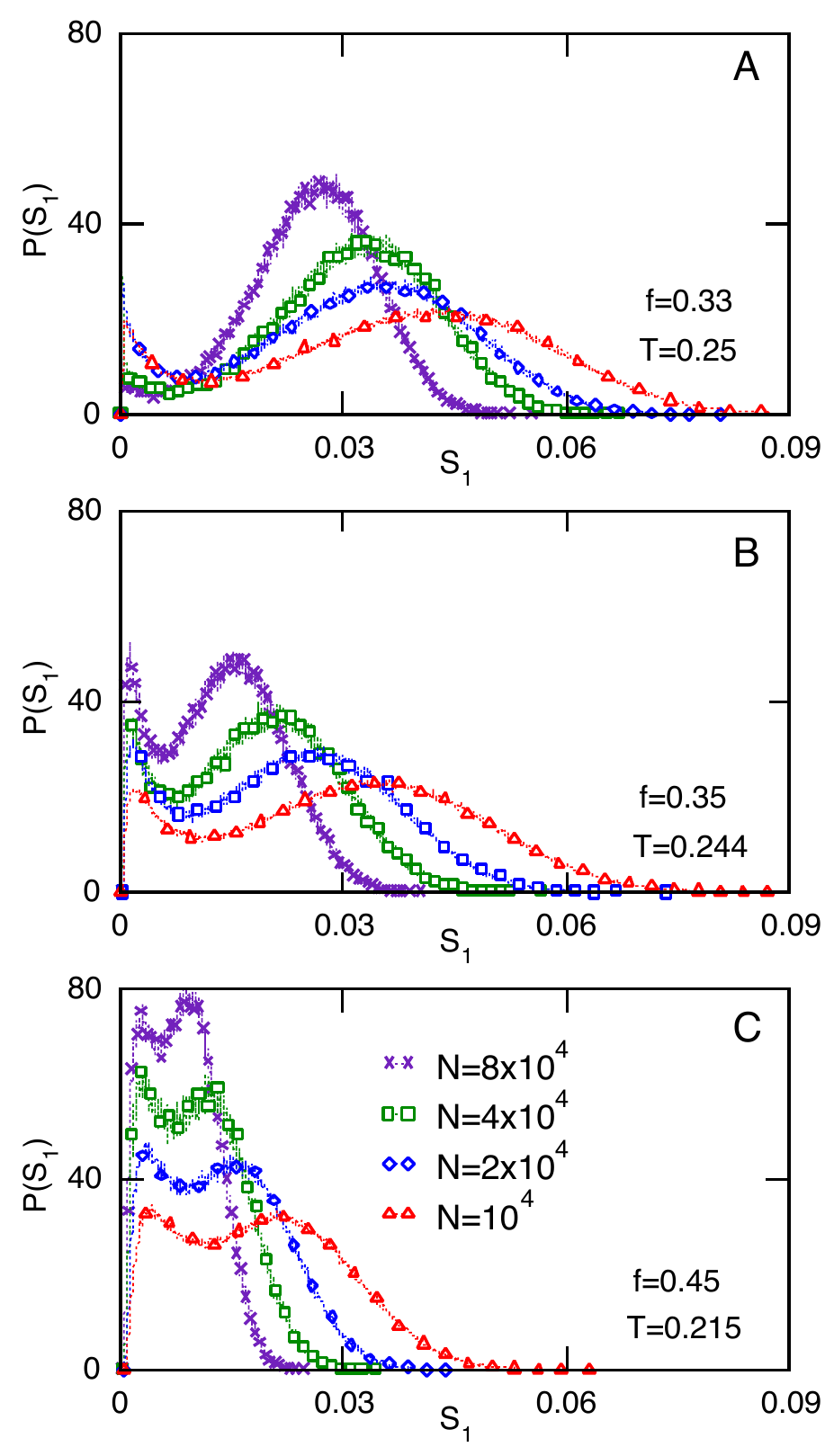}
\caption{\label{fig4} (Color online) Order parameter histograms for a network with  $\langle k \rangle=30$, $\pi=0.6$  for three  $f$ values (denoted in the legend) close to the tricritical point. The threshold values were chosen so that both peaks has approximately the same height.  \emph{Panel A:} for larger sizes the two peaks at the transition are progressively better defined, corresponding to a discontinuous transition. \emph{Panels B, C:} At small sizes the transition looks discontinuous, but on going to larger sizes it is clear that the two peaks are fusing into one, thus showing that the transition is actually continuous.
}
\end{center}
\end{figure}

To further assess the behavior close to the tricritical point  we use the order parameter histogram method described in  section \ref{statistics}. Although this method works very well to characterize discontinuous phase transitions, its usage very close to a critical point presents some subtleties because of  a  particular  type of finite size effects. This is illustrated in Fig.\ref{fig4}. Relatively far away from the critical point and close to the first-order transition point, the two-peak structure of the histogram becomes more marked as the system size increases.  In other words, the location of the peaks converge to well defined distinct values and the minimum between them tends to zero. The fact that the minimum goes to zero for $N/to/infty$ corresponds to the existence of two well-defined and distinct phases in the thermodynamic limit.  An example of such behavior (although weak due to the closeness of the tricritical point) is illustrated in Fig.\ref{fig4}a.  On the other hand, close to the tricritical point (but on the continuous side), both maxima and the minimum tend to collapse into a single maximum when $N\to\infty$, as shown in Fig.\ref{fig4}b and  \ref{fig4}c. Such pseudo-first-order  behavior has already been observed in the two dimensional Potts model with $q=4$\cite{Jin2013}. We estimated the tricritical point location as that where the above-described change in finite-size behavior occurs.
%%%%%%%%%%%%%%%%%%%%%%FIGURE 6%%%%%%%%%%%%%%%%%%%%%

\begin{figure}
\begin{center}
\includegraphics[width=0.42\textwidth]{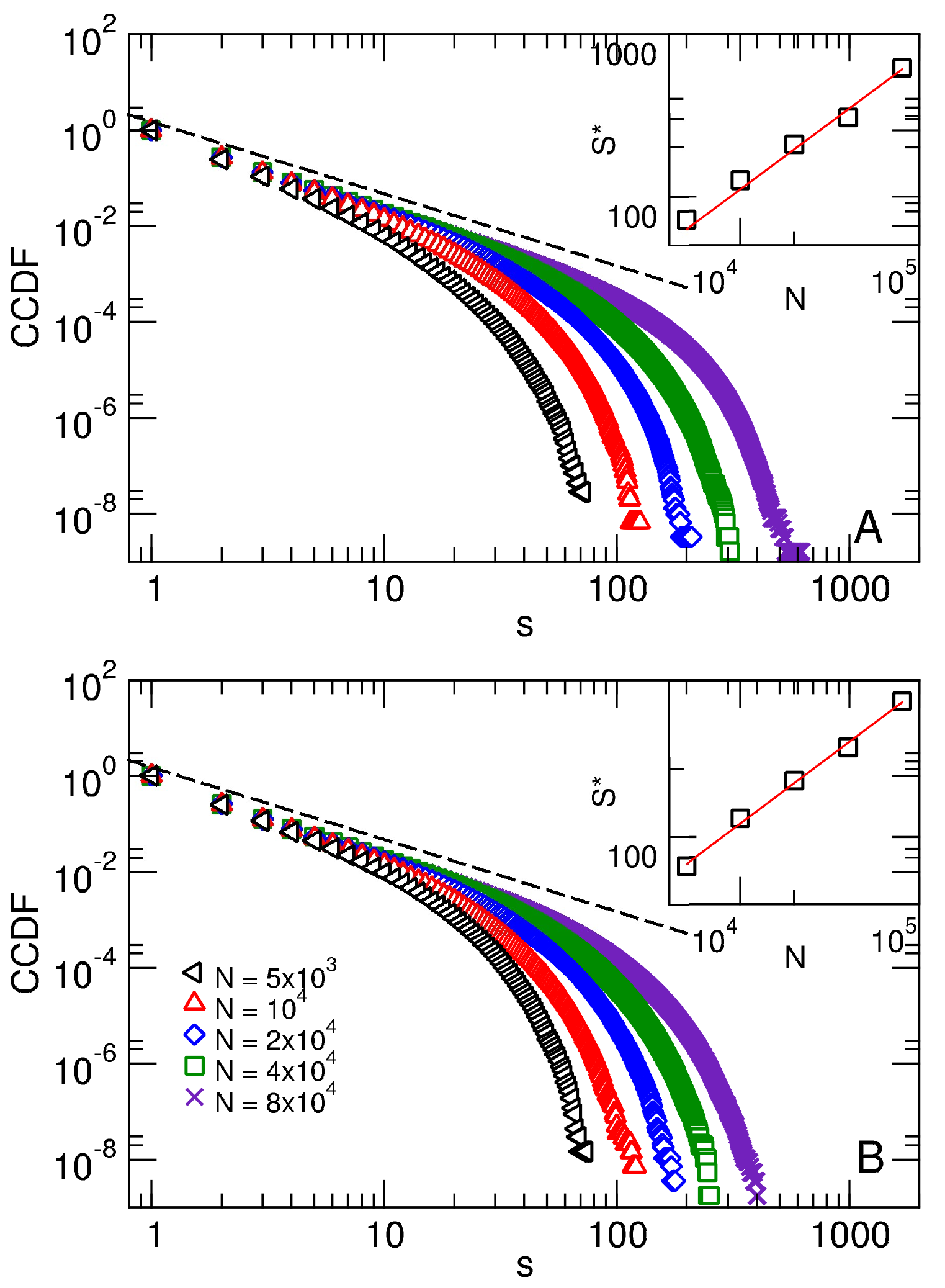}
\caption{\label{fig7} (Color online) CCDF for  $\langle k \rangle=30$, $\pi=0.6$, different values of $N$ and $f$. All the curves were calculated $T=T_c(f)$ (obtained from extrapolation of the $\langle s \rangle$ maxima). Dashed lines correspond to a power law $s^{-\tau+1}$ with $\tau=5/2$. The insets show the cutoff  $S^*$ as a function of $N$. (a) $f=0.5$ with $T_c=0.204\pm 0.001$. Cutoff fitted exponent $d_f/d=0.8\pm 0.2$ (b) $f=0.8$ with $T_c=0.103\pm 0.002$.  Cutoff fitted exponent $d_f/d=0.60\pm 0.06$ }
\end{center}
\end{figure}

Finally, we consider  the universality class of the second order transitions, by estimating the critical exponents $\gamma/\nu d$ and $d_f/d$ from the finite size scaling behavior of the maxima of $\langle s \rangle$  and $\langle S_2 \rangle$. An example is shown in the insets of Fig.\ref{fig5}  for   $\langle k \rangle=30$, $\pi=0.6$ and $f=0.8$. We found that all along the continuous transition line of Fig.\ref{fig6}    the exponents are compatible with the mean-field percolation universality class  $\gamma/\nu d=1/3$ and $d_f/d=2/3$, as observed for $f=0$ and smaller values of $\langle k \rangle$\cite{Zarepour}.  We observed the same behavior even for values of $f$ relatively close to the tricritical point, i.e.\ down to $f=0.359$ (although fluctuations become   larger as we approach the tricritical point, thus increasing the error bars), so we were not able to clearly detect a crossover to a different set of exponents.  To further check the consistency with the mean field percolation universality class we also analyzed the associated behavior of the cluster size distribution $P(s) \equiv N_s/N$ at different  critical values $T=T_c(f)$.  Fig.\ref{fig7} shows that the associated behavior of the cumulative cluster size distribution exhibits the expected behavior  $P_c(s) \equiv \sum'_{s'\geq s} P(s') \sim s^{-(\tau-1)} \exp({-s/S^*})$, with an exponent $\tau \approx 5/2$ and $S^*\propto  \langle S_2 \rangle $(thus satisfying the scaling law $\tau=d/d_f+1$) . 
A similar analysis with similar results was performed for points along the second order line for $k=\langle 16\rangle$ and $\pi=0.6$.

%%%%%%%%%%%%%%%%%%%%%%FIGURE 7%%%%%%%%%%%%%%%%%%%%%

\section{Conclusions}

At first sight, the effects of changing the interaction sign on a fraction of neurons could be interpreted as nothing more than a trivial re-scaling of the excitability control parameter (i.e., the threshold T).  In fact, as shown in Fig. \ref{fig6}, this holds only for relatively small fractions of inhibitory neurons: the discontinuous transition as a function of increasing numbers of inhibitory neurons occurs now for relatively smaller values of T. However, for $f \sim 0.35$ a novel dynamics appears, as the parameter $f$ is increased the line meets the tricritical point and then continues as a second order phase transition.

To interpret its biological relevance, it may be important to recall that the condition for the tricritical point to appear is (besides a large enough fraction of inhibition)  that the network connectivity is very large (i.e. high  $k$). For such highly connected networks, in absence of inhibition there is typically an explosion of highly synchronous bursts in which a very large number of neurons is active, even in response to very small perturbations.  This dynamics, corresponding to a first-order phase transition,  has no behavioral or cognitive value since high synchrony impedes any information processing or storage.   Since the connectivity of cortical neurons is typically in the thousands,  a given fraction of inhibitory neurons can prevent such synchrony.  Is intriguing that the percentage of inhibitions is usually set around 20\%, but probably such quantity cannot be predicted  without accounting for the more complex topology of the real brains compared with the simple W-S network topology studied here.
 
In summary, these results demonstrate that the addition of inhibitory neurons  enriches the dynamical phase diagram observed in Greenberg-Hasting neural models  defined on small world networks \cite{Zarepour}. 
The fraction of inhibitory neurons acts then as an alternative control parameter (in addition to the usual activation threshold) for the dynamical phase transitions between a low activity phase and a percolated, highly active one. Moreover, we observed that the presence of inhibitory neurons allows the emergence of a tricritical point in highly connected networks, i.e., a critical region in parameter space where a second order (i.e., critical) transition hypersurface and a first order (i.e., discontinuous) transition one join smoothly.  We found evidence that, both for large and low values of the connectivity $\langle k \rangle$ the second order surface belongs to the mean-field percolation universality class. On the other hand, we were not able to observe  a crossover to a different set of exponents on approaching the tricritical point for large values of $\langle k \rangle$, due to a large increase in fluctuations, which make an accurate estimation difficult.  This scenario suggests the existence of a tricritical fixed point associated to the tricritical surface (in the sense of renormalization group)  located  far away from the region here analyzed in the parameters space. 

\acknowledgments
This work was partially supported by  CONICET (Argentina) through grants
PIP 11220150100285 and 1122020010106,  by  SeCyT (Universidad Nacional de C\'ordoba, Argentina) and by the NIH (USA) Grant 1U19NS107464-01. JA is a recipient of a Doctoral Fellowship from CONICET (Argentina). This work used Mendieta Cluster from CCAD-UNC, which is part of SNCAD-MinCyT, Argentina.


\begin{thebibliography}{999}

\bibitem{beggs} J.M. Beggs  \& D. Plenz, Neuronal avalanches in neocortical circuits.
 \emph{Journal of Neuroscience}  {\bf 23}, 11167 (2003).

\bibitem{chialvo2010}D.R. Chialvo, Emergent complex neural dynamics. \emph{Nature Physics} {\bf 6,} 744  (2010).


 \bibitem{mora}T. Mora \& W. Bialek, Are biological systems poised at criticality?
 \emph{J. Stat. Phys.}  {\bf 144,} 268 (2011).
 
\bibitem{fraiman}D Fraiman, DR Chialvo. What kind of noise is brain noise: anomalous scaling behavior of the resting brain activity fluctuations, \emph{Frontiers in physiology}{\bf 3}, 307 (2012).

\bibitem{Haimovici2013} A. Haimovici, E. Tagliazucchi, P. Balenzuela, and D. R. Chialvo, \emph{Phys. Rev. Lett.} {\bf 110}, 178101 (2013).

\bibitem{ribeiro} T.L. Ribeiro, S. Yu, D.A. Martin, D. Winkowski, P. Kanold, D.R. Chialvo, D. Plenz, Trial-by-trial variability in cortical responses exhibits scaling in spatial correlations predicted from critical dynamics,
bioRxiv (2020).

\bibitem{tagliazucchi1} E. Tagliazucchi, P. Balenzuela, D. Fraiman and D. R. Chialvo, Criticality in large-scale brain fMRI dynamics unveiled by a novel point process analysis. \emph{Frontiers  in Physiology}, {\bf 3}, 15-15 (2012).

\bibitem{camargo} S. Camargo, D.A. Martin, E.J.A Trejo, A. de Florian, M.A. Nowak, S.A.  Cannas, T.S. Grigera and D.R. Chialvo, Scale-free corr.elations in the dynamics of a small (N~ 10000) cortical network
arXiv preprint arXiv:2206.07797 (2022)



\bibitem{Zarepour} M. Zarepour , J. I. Perotti. O. V. Billoni, D. R. Chialvo and S. A. Cannas,
 \emph{Physical review E} {\bf 100}, 052138 (2019).

%\bibitem{Fox} M.D Fox and M. Raichle, \emph{Nat. Rev. Neurosc.} {\bf 8}, 700 (2007).



\bibitem{Hangmann2008} P. Hagmann, L.  Cammoun, X. Gigandet, R. Meuli, C.J. Honey, V.J. Wedeen, O. Sporns, \emph{PLoS Biol.} {\bf 6}, e159 (2008).
\bibitem{Greenberg-Hastings} J.M. Greenberg and S.P. Hastings, \emph{SIAM (Soc. Ind. Appl. Math.) J. Appl. Math.} {\bf 34}, 515, (1978).

\bibitem{watts-strogatz}  D.J. Watts \& S.H. Strogatz, \emph{Nature} {\bf  393} (6684): 440--2 (1998).

 \bibitem{Barrat} A. Barrat, M Barthelemy \& A . Vespignani, \emph{Dynamical Processes on Complex Networks} (Cambridge Univ, Press (2008).

\bibitem{Margolina} A. Margolina, H.J. Herrmann, and D. Stauffer, \emph{Phys. Lett. A} {\bf 93}, 73 (1982).


\bibitem{Stauffer} D. Stauffer and A. Aharony, \emph{Percolation Theory}. (Taylor and Francis, 2003).

%\bibitem{Tononi} G. Tononi, O. Sporns Edelman, G.   \emph{ Proc. Natl. Acad. Sci. USA} {\bf 91}, 5033--5037 (1994).

\bibitem{Sanchez} M. M. S\' anchez D\'{\i}az, E. J. Aguilar Trejo, D. A. Martin, S. A. Cannas, T. S. Grigera and D. R. Chialvo,  \emph{Physical review E} {\bf 104}, 064309 (2021)

\bibitem{Landau} D. P. Landau and K. Binder, {\em A guide to Monte Carlo simulations in Statistical Physics}, Cambridge University Press (2009).

\bibitem{Jin2013} S. Jin, A. Se, W. Guo and A. W. Sandvik, \emph{Phys. Rev. B} {\bf 87}, 144406 (2013).

\bibitem{Goldenfeld} N. Goldenfeld, {\it Lectures on Phase Transitions and the Renormalization Group}, Frontiers in Physics , Westview Press (1992).

\end{thebibliography}
\end{document}